\documentstyle[12pt,aasms4]{article}

\lefthead{TANIMORI ET AL.}
\righthead{GAMMA RAYS FROM SN1006}	

\begin{document}

\slugcomment{\sl to be submitted to Astrophysical Journal Letters}

\title{DISCOVERY OF TEV GAMMA RAYS FROM SN1006: 
FURTHER EVIDENCE FOR THE SNR ORIGIN OF COSMIC RAYS}
\author{T.~Tanimori\altaffilmark{1}, 
Y.~Hayami\altaffilmark{1}, 
S.~Kamei\altaffilmark{1},
S.~A.~Dazeley\altaffilmark{2}, 
P.G.~Edwards\altaffilmark{3,2},
S.~Gunji\altaffilmark{4},
S.~Hara\altaffilmark{1}, 
T.~Hara\altaffilmark{5},
J.~Holder\altaffilmark{6},
A.~Kawachi\altaffilmark{6},  
T.~Kifune\altaffilmark{6}, 
R.~Kita\altaffilmark{7}, 
T.~Konishi\altaffilmark{8},
A.~Masaike\altaffilmark{9},   
Y.~Matsubara\altaffilmark{10},
T.~Matsuoka\altaffilmark{10},
Y.~Mizumoto\altaffilmark{11}, 
M.~Mori\altaffilmark{6},
M.~Moriya\altaffilmark{1},
H.~Muraishi\altaffilmark{7}, 
Y.~Muraki\altaffilmark{10},
T.~Naito\altaffilmark{12}, 
K.~Nishijima\altaffilmark{13}, 
S.~Oda\altaffilmark{8}, 
S.~Ogio\altaffilmark{1},
J.~R.~Patterson\altaffilmark{2},
M.~D.~Roberts\altaffilmark{6,2}, 
G.~P.~Rowell\altaffilmark{6,2}, 
K.~Sakurazawa\altaffilmark{1},
T.~Sako\altaffilmark{10}, 
Y.~Sato\altaffilmark{6}, 
R.~Susukita\altaffilmark{8,14}, 
A.~Suzuki\altaffilmark{8},
R.~Suzuki\altaffilmark{1},
T.~Tamura\altaffilmark{15}, 
G.~J.~Thornton\altaffilmark{2,6},
S.~Yanagita\altaffilmark{7},
T.~Yoshida\altaffilmark{7}, 
and T.~Yoshikoshi\altaffilmark{6} }

\altaffiltext{1}{Department of Physics, Tokyo Institute of Technology, 
 Meguro-ku, Tokyo 152, Japan}
\altaffiltext{2}{Department of Physics and Mathematical Physics, 
 University of Adelaide, South Australia 5005, Australia}
\altaffiltext{3}{Institute of Space and Astronautical Science, Sagamihara, 
 Kanagawa 229, Japan}
\altaffiltext{4}{Department of Physics,Yamagata University,Yamagata, Yamagata 
990, Japan}
\altaffiltext{5}{Faculty of Management Information, Yamanashi Gakuin 
 University, Kofu, Yamanashi 400, Japan}
\altaffiltext{6}{Institute for Cosmic Ray Research, University of Tokyo, 
  Tanashi, Tokyo 188, Japan}
\altaffiltext{7}{Faculty of Science, Ibaraki University, Mito,
 Ibaraki 310, Japan}
\altaffiltext{8}{Department of Physics, Kobe University, Kobe,
 Hyogo 637, Japan}
\altaffiltext{9}{Department of Physics, Kyoto University, Kyoto,
 Kyoto 606-01, Japan}
\altaffiltext{10}{Solar-Terrestrial Environment Laboratory, Nagoya University,
 Nagoya, Aichi 464, Japan}
\altaffiltext{11}{National Astronomical Observatory of Japan, Mitaka,
 Tokyo 181, Japan}
\altaffiltext{12}{Department of Earth and Planetary Physics, 
 University of Tokyo, Bunkyo-ku, Tokyo 113, Japan}
\altaffiltext{13}{Department of Physics, 
 Tokai University, Hiratsuka, Kanagawa 259-12,Japan}
\altaffiltext{14}{Institute of Physical and Chemical Research, Wako, 
Saitama 351-01, Japan}
\altaffiltext{15}{Faculty of Engineering, Kanagawa University, 
 Yokohama, Kanagawa 221, Japan}

\authoremail{tanimori@hp.phys.titech.ac.jp}
 
\begin{abstract}
This paper reports the first discovery 
of TeV gamma-ray emission from a supernova remnant 
made with the CANGAROO 3.8 m Telescope.
TeV gamma rays were detected at the sky position 
and extension coincident with  the north-east (NE) rim 
of shell-type Supernova remnant (SNR) SN1006 (Type Ia).  
SN1006 has been a most likely  candidate for an  extended TeV Gamma-ray 
source,
since  the clear synchrotron X-ray 
emission from the rims was 
recently observed by ASCA (Koyama et al. 1995), 
which is a strong
evidence of the existence of very high energy electrons up to hundreds of
TeV in the SNR.
The observed TeV gamma-ray flux was
$ (2.4\pm 0.5(statistical) \pm 0.7(systematic)) \times 10^{-12}$ cm$^{-2}$ 
s$^{-1}$ 
($\ge 3.0\pm 0.9$ TeV) and
$ (4.6\pm 0.6 \pm 1.4) \times 10^{-12}$ cm$^{-2}$ s$^{-1}$ ($\ge 1.7\pm 0.5$ 
TeV) 
from the  1996 and 1997 observations, respectively.
Also we set an  upper limit on the TeV gamma-ray emission
from the SW rim, estimated to be 
$ 1.1 \times 10^{-12}$ cm$^{-2}$ s$^{-1}$ ($\ge 1.7\pm 0.5$ TeV, 95\% CL)
in the 1997 data.

The TeV gamma rays can be attributed to the 2.7 K 
cosmic background photons up-scattered by electrons  
of energies up to  about 10$^{14}$ eV 
by the inverse Compton (IC) process.  
The observed flux of the TeV gamma rays, together with that of 
the non-thermal X-rays, gives  firm constraints
on the acceleration process in the SNR shell;
a magnetic field of $6.5\pm2$ $\mu$G is inferred  from  
both the synchrotron X-rays and inverse Compton 
TeV gamma-rays, which gives entirely consistent mechanisms 
that electrons of energies up to 10$^{14}$ eV are 
produced via the shock acceleration in SN1006.

\end{abstract} 

\section{ Introduction }

The origin of high-energy cosmic-rays remains unclear
in spite of the long history of cosmic ray study.
While supernova remnants (SNRs) are the favoured site for cosmic
rays up to 10$^{16}$\,eV,
as they satisfy the required energy input
rate to the galaxy and have a size comparable to the Larmor radius of 
the highest energy particles,
direct supporting evidence is sparse.
Recently detections of GeV gamma-rays
from the regions near several SNRs have been reported by EGRET (Esposito et 
al.\ 1996).
Those results might be evidences of 
the assumed SNR origin of cosmic rays.
However recent observations by the Whipple group of six SNRs, 
including three of EGRET sources,
have given upper limits at hundreds of GeV energy region 
(Buckley et al.\  1998) which are below 
the expected flux from shock acceleration theory
(Drury, Aharonian, \& V\"olk 1994; Naito \& Takahara 1994),
unless cutoffs in the particle spectrum occur
(Sturner et al. 1997; Gaisser, Protheroe, and Stanev 1997).

On the other hand, intense non-thermal X-ray emission 
from the rims of Type Ia SNR SN1006 (G327.6+14.6) has been 
observed by ASCA (Koyama et al.\ 1995), and ROSAT (Willingale et al. 1996),
which is considered,
by attributing the emission to synchrotron radiation, 
to be strong evidence of an existence of high energy 
electrons up to $\sim$100 TeV.
SN1006 is a typical shell-type SNR
which has no apparent central engine for high energy particles
such as a neutron star or a black hole.
Nevertheless, the existence of very high energy particles in a type Ia SNR
is widely accepted
from shock acceleration theory (Blandford \& Eichler 1987; 
Jones \& Ellison 1991).
If so, TeV gamma rays
would also be expected from inverse Compton scattering (IC) of low
energy photons (mostly attributable to the 2.7 K 
cosmic background photons) by these electrons.
By assuming a value for the  magnetic field strength ($B$)
in the emission region of the SNR,
several theorists
( Pohl 1996; Mastichiadis 1996; Mastichiadis \& de Jager 1996;
Yoshida \& Yanagita 1997)
calculated the expected spectra of TeV gamma rays  
using the observed radio/X-ray spectra.
An observation of TeV gamma rays would thus provide not only further direct
evidence of the existence of very high energy electrons
but also  other important information such as the strength of 
the magnetic field and the diffusion coefficient of the shock acceleration.

With this motivation, SN1006 was observed by the CANGAROO
imaging air \v Cerenkov telescope in March and June 1996, 
and  March and April 1997.

\section{Observations}

The observations were made with the 3.8m diameter
\v Cerenkov imaging telescope of the CANGAROO 
Collaboration (Patterson \& Kifune 1992; Hara et al. 1993) 
near Woomera, South Australia (136$^{\circ}$47' E and31$^{\circ}$06' S).
The 3.8 m alt-azimuth mounted telescope 
had a  $\sim$3 TeV threshold for detecting gamma rays 
near 70$^{\circ}$ elevation in the 1996 observations.
The 3.8m mirror was recoated in 1996 October, and its reflectivity
improved from 45\% to more than 80\%,
decreasing the threshold energy by about a factor of two.
A multi-pixel camera consisting
of 256 square photomultiplier tubes, arranged in an array of 0$^{\circ}$.18 
steps,
has a total field of view (FOV) of about 3$^{\circ}$ (Hara et al. 1993).

SN1006 was observed for 28 hours (on-source) and 18 hours 
(off-source) in April and June 1996.
Both of the North East (NE) and the South West (SW) rims
were located within the FOV.
By monitoring the single counting rate in each phototube,
we were able to track the passage of the star (magnitude 3.13) within the FOV, 
and the pointing of the telescope was monitored
to an accuracy of 0$^{\circ}$.02 using the trajectory of this star.
In March and April 1997, in an effort to confirm the 1996 result,
we made additional observations of 34 hours (on source)
and 29 hours (off-source) with the same tracking as in June 1996.

\section{Analysis and Results}

The imaging analysis of the data
is based on the usual parameterization of 
the elongated shape of the \v{C}erenkov
light image as: ``width'',``length'', ``distance'' (location),
``conc'' (shape), and the image orientation angle $\alpha$
(Hillas 1985; Weekes et al.\ 1989).
In the $\alpha$ distribution of the events selected by the imaging
analysis,
the peak appearing around the origin $(\alpha\le 15\arcdeg)$
in the on-source data
is attributed to $\gamma$-rays  from the target position, and 
the number of background events under the peak 
was estimated from the flat region of the $\alpha$ distribution
(30\arcdeg--90\arcdeg) in the on-source data.
Here off-source data were used to verify the non-existence of 
any peculiar structure
in the $\alpha$ plot around the origin not due to  gamma-ray events.  
The application of this technique to data recorded with
the CANGAROO telescope has, to date, resulted in the detection of
TeV gamma rays from PSR1706-44 (Kifune et al.\ 1995) and the nebula
surrounding the
Crab (Tanimori et al.\ 1994; Tanimori et al.\ 1998).
From the results on the previously observed objects, 
the position of a gamma-ray point source
can be determined to an accuracy of 0.1$^{\circ}$. 
The point spread function (PSF) of the CANGAROO telescope 
is estimated to have a standard deviation of 0$^{\circ}$.18 
when fitted with a Gaussian function.

The hard X-ray profile
of the NE rim observed by ASCA
suggests that the TeV gamma rays may emanate from an extended area
over a few times the PSF of the CANGAROO telescope
(several tenths of a degree in extent).
In order to search the emission region  of TeV gamma rays in SN1006, 
significances of peaked events with  $\alpha \le 15^{\circ}$  were
calculated at all grid-points in 0$^{\circ}$.09 steps in the FOV which
is half of the standard deviation of the PSF.
The source point in the NE rim giving the most significant
$\alpha$ peak ($\alpha  \le 15^{\circ}$)
was found  at  the maximum flux point
in the 2--10\,keV band of the ASCA data.
The $\alpha$ plot of the selected gamma-ray-like events at the X-ray 
maximum flux point is  shown in Fig.\ref{fig:f1}a.
Clear peaks due to an excess of gamma-ray events are seen
at $\alpha \sim$ 0$^{\circ}$
for the on-source data of April and June 
but not for the off-source data.
At this X-ray flux maximum point, 
the statistical significances of these peaks are 
estimated to be 3.0$\sigma$ in April,
4.7$\sigma$ in June, and 5.3$\sigma$ in total, 
using the definition mentioned above. 
The resulting contour map of significances is shown in Fig.\ref{fig:f2}a,
in which the contours of the hard X-ray flux 
and the maximum flux point
in the 2--10\,keV band of the ASCA data
also are overlaid as solid bold-lines and marked by a cross,
respectively.
The region showing significant TeV gamma-ray emission
extends along the ridge of the NE rim
over the PSF of the telescope,
and matches the X-ray image  fairly well.

In March and April 1997, in order to confirm the 1996 result,
we made additional observations.
Figure \ref{fig:f1}b shows the $\alpha$ distribution of gamma-ray like events
selected by the same procedure as used for the 1996 data.
A clear peak ($\alpha \le$15$^{\circ}$) was observed again with the 
significance
of 7.7$\sigma$ at the maximum hard X-ray flux point of ASCA.
The improvement of the detection significance was
due to the twice increase of the reflectivity of the mirror. 
Thus the TeV gamma-ray emission from the NE rim of SN1006 has been
confirmed (Tanimori et al.\ 1997).
Figure \ref{fig:f2}b shows the contour map of significances.
The TeV gamma-ray emission region also looks elongated 
along the ridge of the NE rim, 
although the profile of 1997 data
is not as extend as that from the  1996 data.
In order to verify whether the emission region is extended or not, 
the profiles of the NE rim in 1996 and 1997 data were fit
using a superposition of two PSFs located
along the ridge of the NE rim and also using a single PSF.
Although the 1996 result favours the fit using a superposition
of two PSFs against a single PSF,
the 1997 data, with  better statistics, does not show
significant improvement for a superposition of  two PSFs.
From these significance maps we therefore can not claim the extent of the TeV
gamma-ray emission region quantitatively.
Further study is required.   

\placefigure{fig:f1}
\placefigure{fig:f2}

The threshold energy for the observed gamma rays
was determined from Monte Carlo simulations, as
the maximum of the product of the differential flux times
the effective collecting area; 
the latter is a function of gamma-ray energy.
Compared the 1997 result to the 1996 result,
the threshold energy was decreased about a factor of two, and 
the number of the detected gamma-like events was
increased twice.
This indicates a differential photon spectral index 
$d\log N(h\nu)/d\log(h\nu)$ of --2 around a few TeV ($N(h\nu)$ in photons
c$m^{-2}$s$^{-1}$TeV$^{-1}$).
Also the $\alpha$ peak due to gamma-ray like events can be seen
even in about ten times higher energy region than 
the threshold energy in the 1997 data,
which means that there exists little effect of the spectral cut-off in 
the observed energy region. 
Therefore the differential spectral index was simply assumed to be $-2$
in calculating  the integral fluxes of our data.
Results of the Monte Carlo simulation
give a threshold gamma-ray energy of 3$\pm 0.9$ TeV
and an effective area of approximately
6.6$\times$ 10$^8$ cm$^2$ for the 1996 observation of SN1006.
Approximating the emission as coming from a single point source
at the maximum flux  point in the NE rim,
the integral gamma-ray flux for the 1996 observations was calculated
to be $ (2.4\pm 0.5 \pm 0.7) \times 10^{-12}$ cm$^{-2}$ s$^{-1}$ 
($\ge 3.0\pm 0.9$ TeV).
The threshold energy for the 1997 observations
was also estimated to be $1.7\pm 0.5$ TeV.
Using similar approximations to those above,
the gamma-ray integral flux 
for the 1997 observation was calculated to be
$ (4.6\pm 0.6 \pm 1.4) \times 10^{-12}$ cm$^{-2}$ s$^{-1}$ ($\ge 1.7\pm 0.5$ 
TeV).
In those fluxes  the first and second errors are statistical and
systematic respectively.
The systematic error  mainly arises
from the uncertainty of the absolute threshold  energy.
A larger flux would be obtained 
if the  emission extends wider than the PSF
of our detection.
Also the systematic error due to the assumed differential spectrum
was evaluated:
considering the uncertainties in shock acceleration models
such as  nonlinear effects and energy cut off in the electron spectrum
(Jones \& Ellison 1991), 
we varied the spectral index  from --1.2 to --4.0.
In this range, 
the integral flux changes by about --2\% (--1.2) to +20\% (--4.0), which is
relatively smaller than that due to 
the uncertainty of the absolute threshold  energy.

No significant excess is evident in Fig. \ref{fig:f2}a near the position of 
the maximum X-ray flux from the SW rim.
The X-ray  observation by ASCA  indicates that
the integral flux of hard X-rays ($\ge$ 2 keV) in the NE rim
occupies $\sim$60\% of the whole hard X-ray flux emitted from
SN1006 (Ozaki 1997).
The current analysis method of using the $\alpha$ distribution
has difficulty in separating two emission regions as close as 0$^{\circ}$.6
of the  NE and SW rims.
The weaker emission might be hindered by the stronger one from
the NE rim.
Thus, we set an  upper limit on the TeV gamma-ray emission
from the SW rim, estimated to be 
$ 1.1 \times 10^{-12}$ cm$^{-2}$ s$^{-1}$ ($\ge 1.7\pm 0.5$ TeV, 95\% CL)
from the $\alpha$ distribution in the 1997 data 
at the position of the maximum ASCA flux in the SW rim.

\section{Discussion}

This detection of TeV gamma rays  from SN1006
presents a convincing confirmation of 
the shock acceleration mechanism for very high energy particles up to
$\sim$100 TeV in a SNR.
The TeV gamma ray emission region is observed to 
likely extend over $\sim$30 arcmin along the ridge of the NE rim.

From the non-thermal X-ray observation, 
the detected TeV gamma rays are 
readily presumed to be generated by IC scattering 
of very high energy electrons
on $2.7$K cosmic background photons.
All of the calculated fluxes of TeV gamma rays based on 
those assumptions are  consistent
with the TeV gamma-ray fluxes obtained 
by assuming that the  magnetic field strength ($B$)
in the emission region of the SNR is around 10$\mu$G.
One model (Yoshida \& Yanagita 1997)
calculates the expected TeV gamma-ray spectrum
as a function of the strength of the magnetic field
assuming a power law with exponential cutoff energy spectrum
for electrons where the value of necessary parameters are
determined by fitting the observed radio and X-ray synchrotron
emissions.
The observed fluxes of TeV gamma-rays ($\ge1.7$ TeV and $\ge3$ TeV)
fit well if we take $B= 6.5\pm 2\mu$G in the model as shown in 
Fig.\ref{fig:f3}.

\placefigure{fig:f3}

The other candidate of the production mechanism 
of TeV gamma-ray emission is a
decay of neutral pions induced by high energy protons accelerated 
in the SNR.
However, we can neglect the flux from the $\pi^{0}$ decay due to following
arguments.
Since SN1006 (G327.6+14.6) is located above the galactic plane,
the matter density at the shock is low 
($\sim$0.4 cm$^{-3}$: Willingale et al. 1996)
so that the expected flux will be
about a factor of ten less than the observed flux.
The upper limit for GeV gamma-ray emission  from the EGRET archive data
is also consistent with the IC model.
Thus, the detected gamma-rays are likely to be explained
by IC radiation from electrons, and
our result testifies to the existence of the very high energy
electrons of more than several times 10 TeV in SN1006.
The highest energy of non-thermal electrons can be estimated
from the turning point in the synchrotron spectrum and the resultant magnetic 
fields.
Although the turning energy of SN1006
is not yet precisely determined in recent observations,
a 1 keV photon from synchrotron radiation
corresponds to an electron energy of $\sim 60$ TeV
for $B= 6.5 \mu$G.
These values of highest energy and field strength,
and 1000 years of life time for SN1006
are almost consistent with shock acceleration theory.
Observations of some other bands and evolutional theories of the SNR
are required to confirm the highest accelerated energy more precisely
(Sturner et al. 1997; Gaisser, Protheroe, and Stanev 1997).

The observed concentration of hard X-rays and gamma-rays emissions
into the rims in SN1006
also suggests the  possibility that
the relation between the direction of the magnetic field and the shock front
may determine the efficiency of particle acceleration.
Reynolds (Reynolds 1996) pointed out that
the magnetic field in the upstream of the shock of the rims in SN1006
is likely to be parallel to the shock front,
which may show the predominance of highly oblique shocks,
where the efficiency of shock acceleration
is improved (Jokipii 1987; Naito \& Takahara 1995).

Searches for TeV gamma-ray emission from six SNRs
in the northern hemisphere
have been carried out
using a large imaging air \v Cerenkov telescope
by the Whipple group (Buckley et al.\ 1998),
and turned out unsuccessfully.
SN1006 is so far the only SNR in which the existence of very high energy 
electrons up to
$\sim$ 100 TeV  is suspected from X-ray data. 
Recently non-thermal hard X-ray emissions from several
SNRs have been observed
(Koyama et al.\ 1997; Keohane et al.\ 1997; Allen et al.\ 1997),
implying the existence of electrons up to
a few tens of TeV.  
It is clear that more efforts for detecting  TeV gamma rays
from SNRs are necessary to understand 
the shock acceleration mechanism in more detail and therefore the origin of 
cosmic rays.

\acknowledgments

We thank Prof.\ K.~Koyama, Prof.\ J.~Nishimura, Dr.\ M.~Ozaki, 
and Dr.\ O.~de Jager 
for their fruitful discussions.
This work is supported by a Grant-in-Aid in Scientific Research
of the Japan Ministry of Education, Science, Sports  and Culture,
the Australian Research Council and International Science and
Technology Program and the Sumitomo Foundation.
The receipt of JSPS Research Fellowships (P.G.E., T.N., M.D.R., K.S.,
G.J.T.\ and T.~Yoshikoshi ) is also acknowledged.

\clearpage

\figcaption[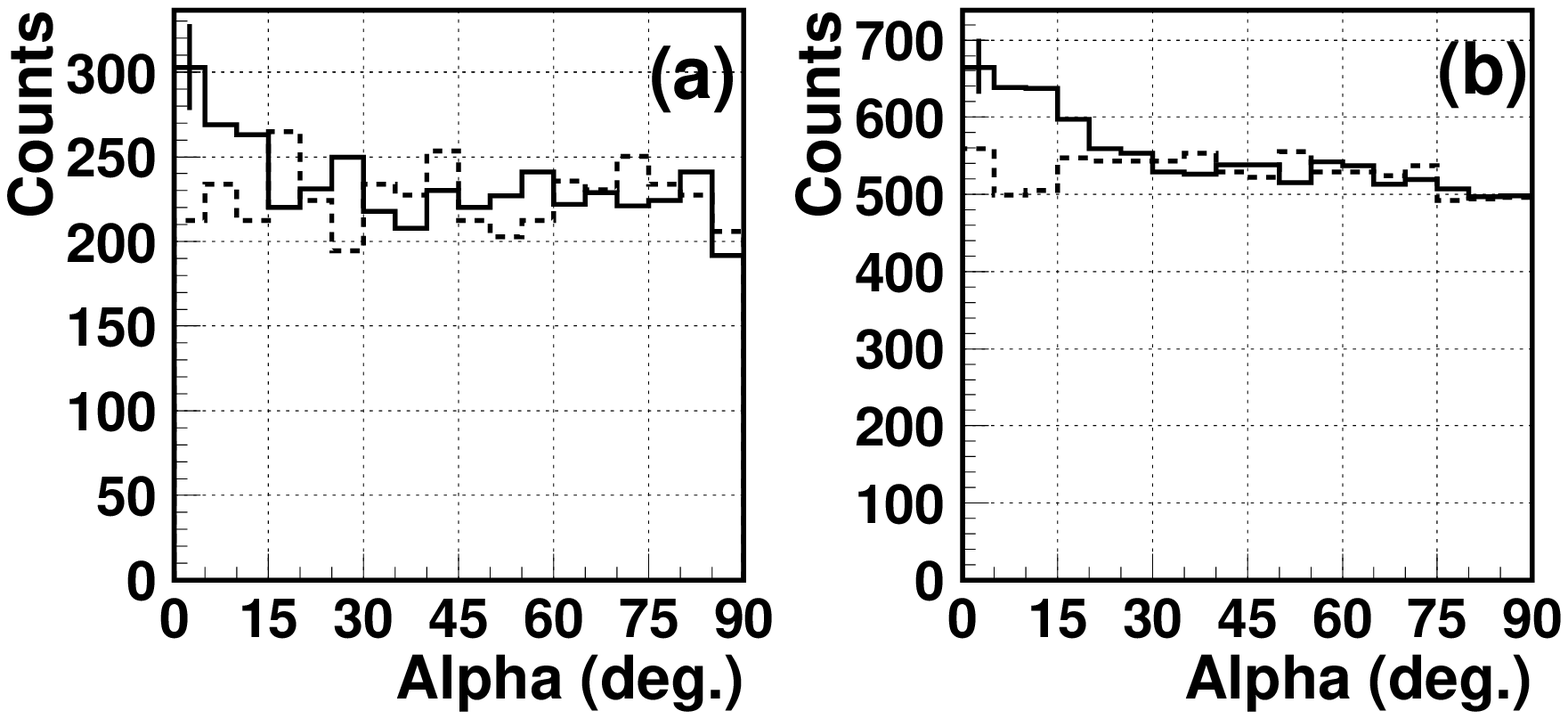]{
(a) The number of observed events as a function of the
orientation angle $\alpha$  for the 1996 data at the maximum flux point of 
hard X-rays  
where on- and off-source data are indicated by the solid and dotted lines, 
respectively.
(b) The same $\alpha$ plot for 1997 data, 
where the on- and off-source data are indicated 
by the solid and dotted lines, respectively. 
Plots of all off-source data are normalized to those of on-source
data by the exposure times.
\label{fig:f1}}

\figcaption[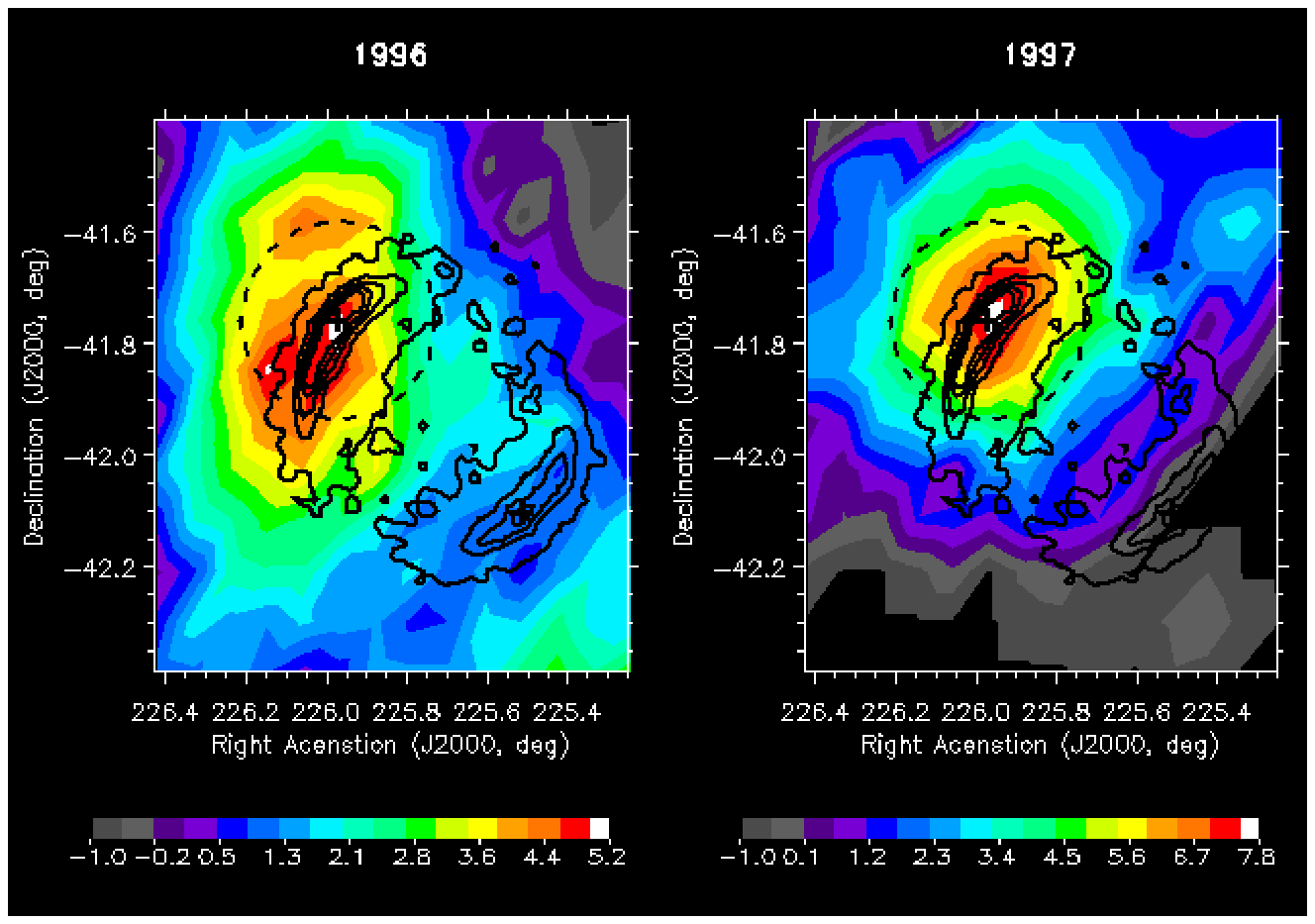]{
(a) The contour map of statistical significance
for various positions in the sky around SN1006.
Also the maximum flux point in the 2--10\,keV band of the ASCA data
is marked by a cross.
The dashed circle is the area of the point spread function
of the CANGAROO telescope within which the significance is larger 
than half the maximum value.
(b) The same contour map obtained from the 1997 data. 
\label{fig:f2}}

\figcaption[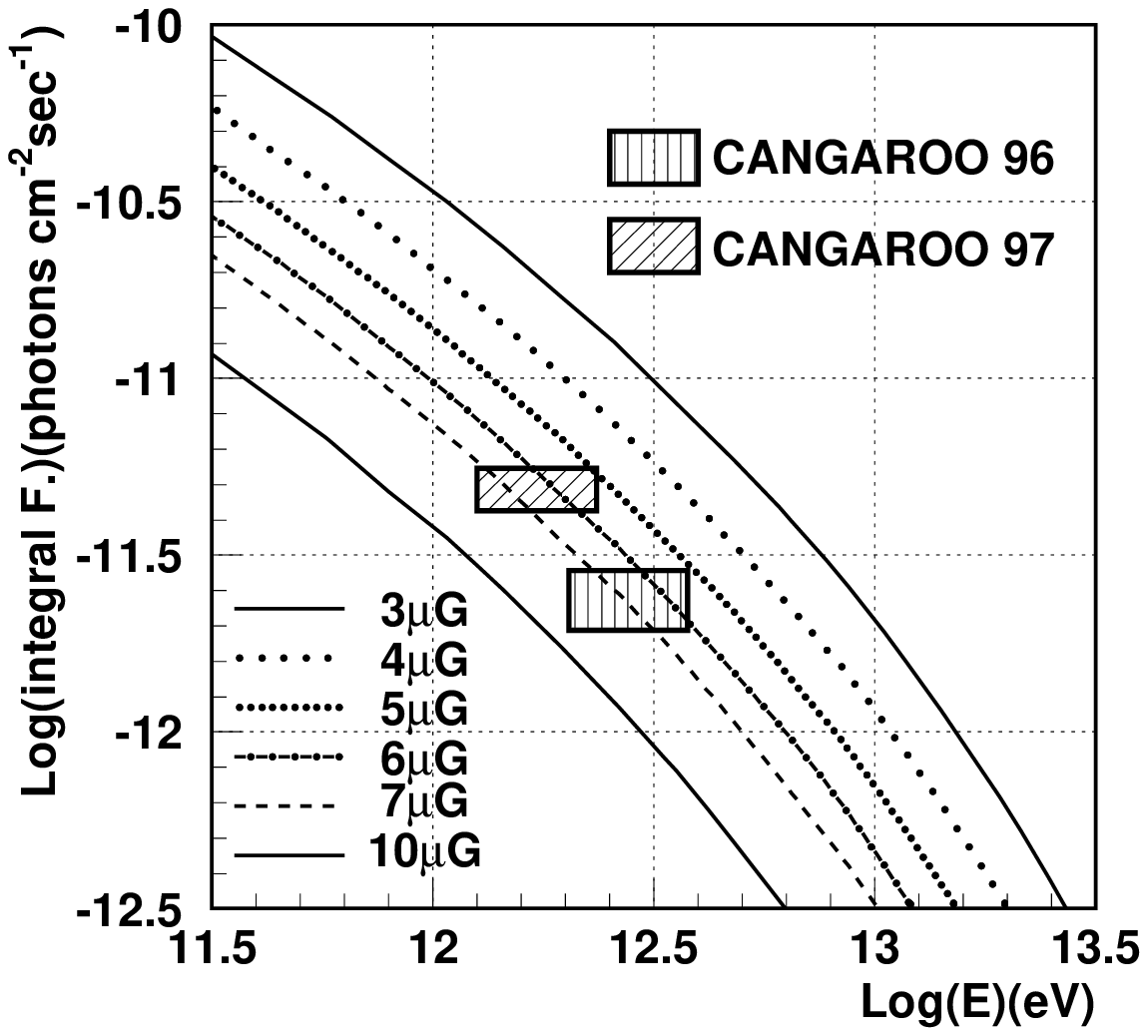]{
The expected integral spectra of TeV gamma rays for several values of
the  magnetic field strength {\it B} calculated by  Yoshida \& Yanagita (1997)
are compared to the observed TeV gamma ray fluxes from 
the 1996 and 1997 data (in the rectangular boxes).
The best estimation of {it B} from the experimental data is $6.5 \pm 2 \mu$G.
\label{fig:f3}}

\clearpage
\begin{figure}
\plotone{snfig1.eps}
\end{figure}

\clearpage
\begin{figure}
\plotone{snfig2.eps}
\end{figure}

\clearpage
\begin{figure}
\plotone{snfig3.eps}
\end{figure}

\end{document}